\documentclass[11pt]{llncs}
\usepackage[T1]{fontenc}
\usepackage{graphicx}
\usepackage{amsmath}
\usepackage{amssymb}
\usepackage{xspace}
\usepackage{todonotes}
\usepackage{tikz}
\usetikzlibrary{arrows}
\usetikzlibrary{shapes}
\usepackage{algorithm}
\usepackage[noend]{algorithmic} 
\usepackage{wrapfig}
\usepackage{multirow}
\usepackage{hyperref}

\usepackage{geometry} 

\newcommand{\block}[1]{~\\[-3mm] \noindent {\textsc{#1.}}~}
\DeclareFontFamily{U}{mathb}{\hyphenchar\font45}
\DeclareFontShape{U}{mathb}{m}{n}{
      <5> <6> <7> <8> <9> <10> gen * mathb
      <10.95> mathb10 <12> <14.4> <17.28> <20.74> <24.88> mathb12
      }{}
\DeclareSymbolFont{mathb}{U}{mathb}{m}{n}
\DeclareMathSymbol{\dotdiv}{2}{mathb}{"01}
\DeclareUnicodeCharacter{2212}{\ensuremath{-}}

\newcommand{\miniscule}{\@setfontsize\miniscule{3}{3.5}}

\newcommand{\Look}{{\tt Look}\xspace}
\newcommand{\Compute}{{\tt Compute}\xspace}
\newcommand{\Move}{{\tt Move}\xspace}
\newcommand{\LCM}{{\tt LCM}\xspace}

\newcommand{\pre}{\mathtt{pre}}

\newcommand{\oblot}{\textsc{Oblot}\xspace}

\newcommand{\LSS}{\mathit{LSS}}
\newcommand{\Ex}{\mathbb{E}} 
\newcommand{\A}{\mathcal{A}} 
\newcommand{\cA}{\mathcal{A}} 
\usepackage{xcolor}

\newcommand{\h}{\mathit{H}}

\newcommand{\MR}{\mathit{MR}}

\newcommand{\GMV}{\mathrm{\ensuremath{GMV}}}  

\newcommand{\algo}{\mathcal{A}} 

\newcommand{\true}{\mathtt{true}}

\newcommand{\RP}{\mathit{RP}}
\newcommand{\RC}{\mathit{RC}}
\newcommand{\FV}{\mathit{FV}}
\newcommand{\ERdown}{\mathit{ER_{down}}}
\newcommand{\ERup}{\mathit{ER_{up}}}

\newcommand{\specialpath}{special-path\xspace}
\newcommand{\specialpaths}{special-paths\xspace}
\newcommand{\specialsubpath}{special-subpath\xspace}

\newcommand{\FO}{\mathit{fos}}

\newcommand{\MoveAlongSpecialpath}{\textsf{MoveAlongSpecial-Path}}
\newcommand{\MoveAlongColumns}{\textsf{MoveAlongColumns}}

\makeatletter
\newcommand{\checkheight}[1]{%
  \par \penalty-100\begingroup%
  \setbox8=\hbox{#1}%
  \setlength{\dimen@}{\ht8}%
  \dimen@ii\pagegoal \advance\dimen@ii-\pagetotal
  \ifdim \dimen@>\dimen@ii
    \break
  \fi\endgroup}
\makeatother

\begin{document}
%
\title{%
An optimal algorithm for geodesic mutual visibility on hexagonal grids
}
%
%

\author{%
Sahar {Badri}\inst{1} 
\and
Serafino {Cicerone}\inst{1}
\and
Alessia {Di Fonso}\inst{1}
\and
Gabriele {Di Stefano}\inst{1}
}
\authorrunning{A. Di Fonso et al.}
%
\institute{Department of Information Engineering, Computer Science and Mathematics (DISIM), University of L'Aquila, Italy
\\
\email{alessia.difonso@univaq.it}, 
\email{sahar.badri@graduate.univaq.it}, 
\email{serafino.cicerone@univaq.it}, \email{gabriele.distefano@univaq.it}
\url{}}
%
\maketitle              
%

\begin{abstract}
For a set of robots (or agents) moving in a graph, two properties are highly desirable: confidentiality (i.e., a message between two agents must not pass through any intermediate agent) and efficiency (i.e., messages are delivered through shortest paths). These properties can be obtained if the \textsc{Geodesic Mutual Visibility} ($\GMV$) problem is solved: oblivious robots move along the edges of the graph, without collisions, to occupy some vertices that guarantee they become pairwise geodesic mutually visible. This means that there is a shortest path (i.e., a ``geodesic'') between each pair of robots along which no other robots reside.
In this work, we optimally solve $\GMV$ on finite hexagonal grids $G_k$. This, in turn, requires first solving a graph combinatorial problem, i.e. determining the maximum number of mutually visible vertices in $G_k$.

\end{abstract}
%
%
%
\section{Introduction}
Problems about sets of points in the Euclidean plane and their mutual visibility have been investigated for a long time. For example, in~\cite{Dudeney17} Dudeney posed the famous \emph{no-three-in-line} problem: finding the maximum number of points that can be placed in an $n \times n$ grid such that there are no three points on a line. 

Mutual visibility in graphs for a set of vertices has been recently introduced and studied in~\cite{D22} in terms of the existence of a shortest path between two vertices without a third vertex from the set.
The visibility property is then understood as the absence of ``obstacles'' between the two vertices along the shortest path, which makes them ``visible'' to each other. For example, in communication networks, mutually visible agents can communicate both efficiently (i.e., through shortest paths) and confidentially (i.e., the message does not pass through intermediate agents).

Formally, let $G$ be a connected and undirected graph, and $X\subseteq V(G)$ a subset of the vertices of $G$. Two vertices $x, y \in V(G)$ are \emph{$X$-visible} if there exists a shortest $x,y$-path where no internal vertex belongs to $X$, and $X$ is a \emph{mutual-visibility set} if its vertices are pairwise $X$-visible. Any largest mutual-visibility set of $G$ is called $\mu$-set and its cardinality is the \emph{mutual-visibility number} of $G$ (denoted as $\mu(G)$). 
In~\cite{D22}, it is shown that computing $\mu(G)$ is an NP-complete problem. Still, exact formulae exist for the mutual-visibility number of special graph classes like paths, cycles, blocks, cographs, grids and distance-hereditary graphs~\cite{CICERONE2023LAGOS}. 


Ever since its introduction, this concept of graph-based mutual-visibility has gardened significant interest within the research community. This resulted in a remarkable list of articles~\cite{axenovich2024,Bresar,BujtaKT23,CDDNP23,CICERONE2023LAGOS,variety23,CiceroneDK23,CiceroneDKY23,CiceroneDKY24,ekinci2024,TianK22,Korze2024,kuziak-2023,NavarraP24,roy2024,tian-2023+}. 
These contributions provided new structural and computability results and identified several connections between the mutual-visibility problem and some classical combinatorics topics. For example, there exist relationships with the Zarankiewicz problem~\cite{CiceroneDK23}, the Turán problem~\cite{BujtaKT23,CiceroneDKY24}, and the classical Bollob\'as-Wessel theorem~\cite{Bresar}. There is also a close relationship between the mutual-visibility problem and the general position problem~\cite{ullas-2016,KlavzarNC22,manuel-2018}, also a topic distance-related that has attracted great interest in recent years.

The graph-based notion of mutual-visibility has already been applied in the context of mobile robots operating on discrete environments modeled by graphs where visibility is verified along shortest paths (cf.~\cite{CDDN23b,CDDN23-PMC,CDDN2023-gmv-grids}). In particular, starting from a configuration composed of any number of robots located on distinct vertices of an arbitrary graph, within finite time the robots must reach (if possible), a configuration where they are all mutually visible. This problem is called \textsc{Geodesic Mutual Visibility} problem ($\GMV$, for short) to distinguish it from the well-known \textsc{Mutual Visibility} problem on the Euclidean plane, where two robots are visible if there is not a third robot on the straight line segment between them (e.g., see~\cite{ABKS18,DFCPSV17,DFPSV14}). 
%

\begin{figure}[h]
\centering
\includegraphics[width=0.6\textwidth]{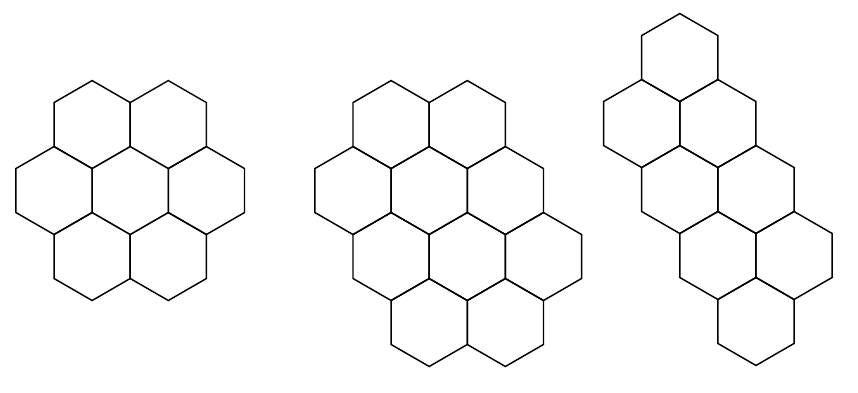}
\caption{Examples of hexagonal $k\times m\times n$ grids: \emph{(left)} a $2\times 2\times 2$ grid; \emph{(center)} a $2\times 2\times 3$ grid; \emph{(right)} a $1\times 2\times 4$ grid.}
\label{fig:grids}
\end{figure}

\block{Results} In this paper, we are interested in solving $\GMV$ for synchronous robots moving on finite hexagonal grids embedded in the plane. 
Such grids are denoted as $G_k$, $k\ge 1$, and correspond to the $k\times k\times k$ grid graph defined in~\cite{hexgrid-wolfram} (cf. Figure~\ref{fig:grids}). Note that, regular grid graphs are a typical discretization of the plane used in swarm robotics 
(for example under the \oblot model~\cite{ABKS18,BAKS20,C21,CDDN23,CiceroneFSN24,DN17,HectorSVT22}, and the programmable matter model~\cite{Navarra-pr-matter,DaymudeRS23,DerakhshandehGS15}).


It is worth noting that any algorithm solving $\GMV$ on a graph $G$ for any possible number of robots must know not only $\mu(G)$ but also some $\mu$-set of $G$ to use it as a pattern defining the final positions that the robots must reach. For this reason, our contribution is twofold. First, from a graph-theoretical point of view, we provide an exact formula for $\mu(G_k)$ and determine a $\mu$-set for $G_k$.
Then, we design an algorithm $\algo$ that solves optimally $\GMV$ in each configuration defined on $G_k$ and composed of synchronous robots endowed with chirality. Algorithm $\algo$ uses the $\mu$-set of $G_k$ as a pattern. We remark that the general \textsc{Pattern Formation} problem (where robots belonging to an initial configuration $C$ are asked to move and form a configuration
$F$ -- i.e., the pattern -- that is provided as input to robots) has been already studied on grid graphs, but all the algorithms proposed so far consider only asymmetric configurations in input, while $\algo$ must solve $\GMV$ even when the initial configuration is symmetric (e.g., see~\cite{BAKS20,CDDN23,GGSS23}). 


\section{Robot model and the addressed problem}
An \oblot system (cf.~\cite{FPS19a}) comprises a set of robots that live and operate in a graph. 
Robots are \textbf{identical} (indistinguishable from their appearance), \textbf{anonymous} (they do not have distinct ids), 
\textbf{autonomous} (they operate without a central control or external supervision), \textbf{homogeneous} (they all execute the same algorithm), \textbf{silent} (they have no means of direct communication of information to other robots), and \textbf{disoriented} (each robot has its own local coordinate system - LCS) but they agree on a cyclic orientation (e.g., clockwise) of the plane, i.e., a common sense of \textbf{chirality} is assumed. A robot can observe the positions (expressed in its LCS) of all the robots. We consider \textbf{synchronous} robots that operate according to the \Look-\Compute-\Move  (\LCM) computational cycle~\cite{FPS19a}:
\begin{itemize}
\item  \Look. The robot obtains a snapshot expressed in its own LCS of the positions of all the other robots. 
\item  \Compute. The robot performs a local computation according to a deterministic algorithm $\cA$ (i.e., the robot executes $\cA$), which is the same for all robots, and the output is a vertex among its neighbors or the one where it resides.
\item  \Move. The robot performs a \emph{nil} movement if the destination is at its current location otherwise it \textbf{instantaneously moves} to the computed neighbor.
\end{itemize}
Robots are \textbf{oblivious} (they have no memory of past events), thus the \Compute phase depends only on the information of the current \Look phase. A data structure containing the information elaborated from the current snapshot represents what is called the \textbf{view} of a robot. Since each robot refers to its own LCS, the view cannot exploit absolute orienteering and it is based on the relative positions of robots. Hence, if symmetries occur (see Section~\ref{sec:notation}), then symmetric robots have the same view. In turn, (i) the algorithm cannot distinguish between symmetric robots -- even when placed in different positions, and (ii) symmetric robots perform the same movements. 

Robots are placed in a simple, undirected, and connected graph $G=(
V,E)$. 
A function $\lambda: V\to \mathbb{N}$ represents the number of robots on each vertex of $G$, and we call $C=(G,\lambda)$ a \textbf{configuration} whenever $\sum_{v\in V} \lambda(v)$ is bounded and greater than zero. A vertex $v\in V$ such that $\lambda(v)> 0$ is said \emph{occupied}, \emph{unoccupied} otherwise. 
We say that a collision occurs on a vertex $v$ if $\lambda(v)>1$.
%

\begin{definition}[$\GMV$ problem]\label{def:GMV}
Let $C=(G,\lambda)$ be any configuration with $\lambda(v)\le 1$ for each $v\in G$. Design a deterministic distributed algorithm working under the \LCM model that, starting from $C$ and in a finite number of computational cycles, moves the robots without collisions until they form a configuration $C'=(G,\lambda')$ where the occupied vertices are in mutual visibility.
\end{definition}

Here we study $\GMV$ in finite hexagonal grids. Note that, given a configuration $C=(G,\lambda)$ with $n$ robots, any algorithm designed to solve $\GMV$ on $C$ must be provided with a strategy that, in a finite time, guides robots to occupy vertices that are in mutual-visibility. This can be obtained by providing the algorithm with a mutual-visible set $X$ of $G$ such that $|X|\ge n$. To the best of our knowledge, computing $\mu(G)$ when $G$ is a finite hexagonal grid is an open problem. 

\section{Mutual visibility on hexagonal grid graphs}\label{sec:muset-Gk}
%


\begin{figure}[t]
\centering
\def\svgwidth{0.75\columnwidth}
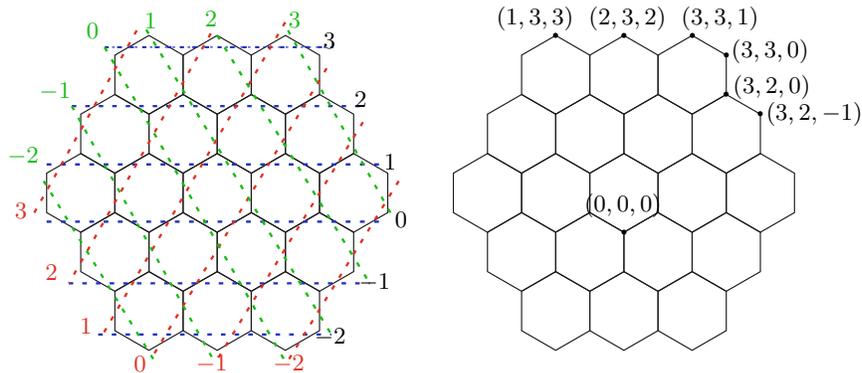
\caption{\textit{%
\emph{(left)} Visualization of $G_3$ with the three families of oriented lines used for the vertex labeling; 
\emph{(right)} Visualization of some vertex labels;} 
}
\label{fig:G_k}
\end{figure}
We consider finite subgraphs of \textbf{hexagonal grid graphs}: $G_1$ is the grid graph of just one hexagon, $G_2$ is obtained by surrounding $G_1$ with a ``crown'' of 6 additional hexagons -- one per side, and, in general, $G_k$ with $k\ge 2$ can be informally built by surrounding $G_{k-1}$ with a crown of $6(k-1)$ additional hexagons (see Figure~\ref{fig:G_k}). 
Formally, $G_k$ corresponds to the $k\times k\times k$  grid graph defined in~\cite{hexgrid-wolfram} and results in a finite graph with $6k^2$ vertices. 
It can be divided into 6 \textbf{sectors} by using the three lines passing through the middle points of parallel edges of the initial sub-grid $G_1$. The \textbf{perimeter} of $G_k$ is formed by all the external vertices and edges of the grid, and it consists of 6 \textbf{sides} formed by a path of length 1 when $k=1$ and of length $2k$ when $k\ge 2$ 
(in these grids, two consecutive sides share one edge). Vertices can be labeled by extending the classical Cartesian coordinates used in the Euclidean plane to three directions. To this end, consider the three families of straight lines parallel to the sides of $G_k$: according to the orientation, 
we call them r-lines (right-oriented lines), l-lines (left-oriented lines) and h-lines (horizontal lines); sometimes, h-lines are also called ``levels''. These lines can be numbered according to the distance from the lowest vertex of the central hexagon,
starting from 0.
(cf. Figure~\ref{fig:G_k}). 
A vertex at the intersection of three lines is labeled by a triple consisting of the numbers assigned to those lines in the order (l-line, h-line, r-line).%
\footnote{Note that this labeling is used only for assessing the mutual visibility of $G_k$ and cannot be later exploited by robots; robots do not have a global reference system.} 
%
%
%
%

\begin{figure}[t]
   \centering
\scalebox{1.2}{ 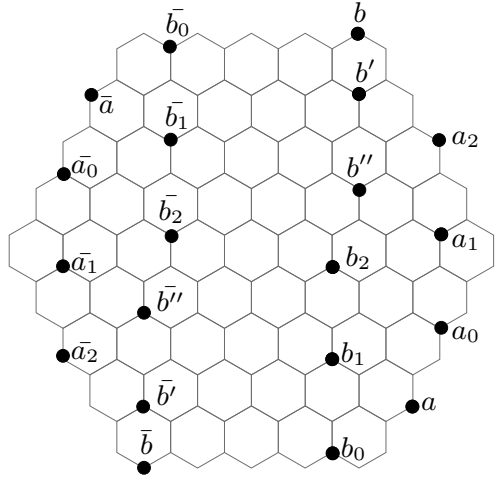}
\caption{Visualization of the $\mu$-set $X_k$ when $k=5$.} 
\label{fig:optG_5}
\end{figure}

 
We now define a $\mu$-set $X_k$ for $G_k$ for each $k\ge 2$. To this end, consider the following subsets (cf Figure~\ref{fig:optG_5}):
\begin{itemize}
\item 
$A_k = \{ a \} \cup \{ a_i | i=0,\ldots,k-3 \}$, where $a=(2,-k+2,-k+1)$ and $a_i =  (3+i,-k+4+2i,-k+1+i)$; 
\item 
$B_k = \{ b, b', b''\} \cup \{ b_i | i=0,\ldots,k-3  \}$, where $b=(k,k,1)$, $b'=(k-1,k-1,0)$, $b''=(k-2,k-3,-1)$ and  $b_i = (i,-k+1+2i,-k+2+i)$; 
\item 
$\bar{A}_k$ ($\bar{B}_k$, resp.) contains the vertices of $G_k$ obtained through a $180^{\circ}$ rotation of the elements of $A_k$ ($B_k$, resp.).
\end{itemize}
Finally, let $X_k = A_k \cup B_k \cup \bar{A}_k \cup \bar{B}_k$. Note that $|A_k|=k-1$, $|B_k|=k+1$, and hence $|X_k|=4k$. The next result states that $\mu(G_k) = 4k$.
\begin{theorem}\label{thm:muset-Gk}
$X_k$ is a $\mu$-set of $G_k$, for each $k\ge 4$. 
\end{theorem}

\begin{proof}
A subgraph $H$ of any graph $G$ is called convex if all shortest paths in $G$ between vertices of $H$ belong to $H$. In~\cite[Lemma~2.1]{D22} it is shown that if $H$ is a convex subgraph of any graph $G$, and $X$ is a mutual-visibility set of $G$,  then $X\cap V(H)$ is a mutual-visibility set of $H$. 
Observe now that each line of $G_k$ is a convex subgraph. Since this line is isomorphic to a path $P$ and $\mu(P)=2$, by~\cite[Lemma~2.8]{D22} we get that any mutual visibility set for $G_k$ share at most 2 vertices with each line. Since the vertices of $G_k$ can be partitioned into $2k$ h-lines, we get $\mu(G_k)\le 4k$. 

To conclude the proof, we show that the set $X_k$, having $4k$ vertices, is indeed a mutual-visibility set for $G_k$. To this aim, we have to show that if $u,v\in A_k\cup B_k\cup \bar{A}
_k\cup \bar{B}_k$, $u\neq v$, then there exists a $u,v$-shortest path not containing any internal vertex in $X_k$. In principle, since both $u$ and $v$ may belong to 4 different subsets, then we should analyze 16 different cases, but due to the symmetry only the following 6 cases occur:

\begin{enumerate}
\item 
Let $u\in A_k$ and  $v\in \bar{B}_k$. Recalling that $A_k=\{a\}\cup \{a_i~|~ 0\le i\le k-3\}$, assume first $u\not = a$. For each vertex $a_i\in A_k$, $0\le i\le k-4$, consider the vertex $c_i$ on the same h-line of $a_i$ at distance 1 from $b_{i+1}$. They are all at a distance $4$ from $a_i$. For the vertex $a_{k-3}$ (the vertex with maximum index $i=k-3$) consider the vertex $c_{k-3}$ on the same line of $a_{k-3}$ and at distance 4 from $a_{k_3}$.
Formally $c_i=(i+1,-k+4+2i,-k+3+i)$. Let $C_k$
the set of all the vertices $c_i$. Vertices in $C_k\cup\bar{B}_k$ are in mutual visibility and
then a shortest $u,v$-path $P$, with $u=a_i$ can be found such that $P$ passes through vertex $c_i$.
When $u= a$ consider the vertex $c=(0,-k+2,-k+2)$ on the same h-line of $a$ and at distance 1 from $b_0$.
Then a shortest $a,v$-path $P$ can be found such that $P$ passes through vertex $c$ for each vertex $v\in \bar{B}_k\setminus \{\bar{b}_0\}$. However, $a$ and $\bar{b}_0$ are in mutual visibility since there is a shortest $a,\bar{b}_0$-path passing through $c_0$.
\item 
We now check the mutual visibility between any vertex in $A_k$ against any vertex in $\bar{A}_k$. Let $\bar{C}_k$ be the set of vertices of $G_k$ obtained through a $180^{\circ}$ rotation of the elements of $C_k$ and let $\bar{c}$ the vertex corresponding to $c$ obtained with the same rotation. Consider two vertices $a_i\in A_k\setminus{a}$ and $\bar{a}_j\in \bar{A}_k\setminus\{\bar{a}\}$, for any pair of indices $i,j \in\{0,\ldots, k-3\}$. Then, a shortest path $P$ between $a_i$ and $\bar{a}_j$ can be found as follows. $P$ starts from $a_i$, passes through $c_i$ on the same h-line of $a_i$, then goes towards $\bar{c}_j$ and finally reaches $\bar{a}_j$. 
Vertex $a$ is in mutual visibility with any vertex $\bar{a}_j\in \bar{A}_k\setminus\{\bar{a}\}$, for any index $j \in\{0,\ldots, k-3\}$. Indeed, a shortest path between $a$ and $\bar{a}_j$ can be found passing on $c$ and $\bar{c}_j$. Similarly for the visibility of vertex $\bar{a}$ with respect to vertices in $A_k$. Finally, $a$ and $\bar{a}$ are in mutual visibility since there is a shortest $a,\bar{a}$-path passing through $c$ and $\bar{c}_0$.
\item 
Let $u,v\in A_k$. These vertices are in mutual visibility since for each pair there is a shortest $u,v$-path $P_{u,v}$ passing on vertices located between $A_k$ and $B_k$ and not in $X_k$. Indeed, $P_{u,v}$ is such that each vertex is at a distance at most 3 from a vertex in $A_k$ and, since the distance between two vertices one in $A_k$ and the others in $B$ is at least 4, no internal vertex of $P_{u,v}$ is in $X_k$. 
\item 
Let $u,v\in B_k$. We can apply here the same arguments used in the previous case. 
\item 
Let $u\in A_k$ and $v\in B_k$. Since, as observed, the distance between any pair of vertices $x,y$ such that $x\in A_k$ and $y\in B_k$ is at least 4, then it is easy to see that there always exists a shortest $u,v$-path passing on vertices located between $A_k$ and $B_k$ not in $X_k$. 
\item 
Let $u\in B_k$ and $v\in \bar{B}_k$. All the shortest $u,v$-paths do not involve vertices in $A_k\cup \bar{A}_k$, and there is at least a shortest $u,v$-paths not involving other vertices in $B_k \cup \bar{B}_k$.
\end{enumerate}
This concludes the proof.
\qed
\end{proof}

\section{Notation and concepts for $\GMV$}\label{sec:notation}
%
%
%
Given $C=(G_k,\lambda)$, we use $R = \{r_1, r_2,\ldots, r_n\}$ to denote the set of robots located on $C$.%
\footnote{We recall that robots are anonymous and such a notation is used only for the sake of presentation, hence no algorithm can take advantage of names of robots.}
The distance between the vertices $u$ and $v$ is denoted by $d(u, v)$, and given $r_i,r_j\in R$, $d(r_i, r_j)$ represents the distance between the vertices in which the robots reside. Finally, $D(r) =\sum_{r_i\in R}{d(r,r_i)}$. 

\block{Symmetric configurations}\label{ssec:symmetries}
As chirality is assumed, rotations are the only possible symmetries in $G_k$. A rotation is defined by the center $c$ of $G_k$ and a minimum angle of rotation $\alpha\in \{60,120,180,360\}$ working as follows: if the configuration is rotated around $c$ by an angle $\alpha$, then a configuration coincident with itself is obtained. The \textbf{symmetricity} of a configuration $C$ is denoted as $\rho(C)$ and corresponds to $360/\alpha$. 
$C$ is \textbf{rotational} if $\rho(C)>1$, and is \textbf{asymmetric} if $\rho(C)=1$.

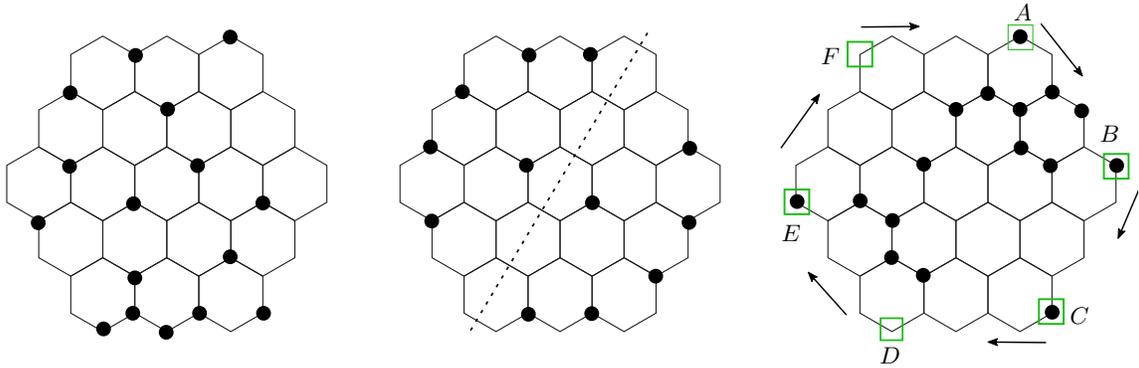
\begin{figure}[t]
   \centering
   \def\svgwidth{1.0\columnwidth}
\begingroup%
  \makeatletter%
  \providecommand\color[2][]{%
    \errmessage{(Inkscape) Color is used for the text in Inkscape, but the package 'color.sty' is not loaded}%
    \renewcommand\color[2][]{}%
  }%
  \providecommand\transparent[1]{%
    \errmessage{(Inkscape) Transparency is used (non-zero) for the text in Inkscape, but the package 'transparent.sty' is not loaded}%
    \renewcommand\transparent[1]{}%
  }%
  \providecommand\rotatebox[2]{#2}%
  \newcommand*\fsize{\dimexpr\f@size pt\relax}%
  \newcommand*\lineheight[1]{\fontsize{\fsize}{#1\fsize}\selectfont}%
  \ifx\svgwidth\undefined%
    \setlength{\unitlength}{317.92516934bp}%
    \ifx\svgscale\undefined%
      \relax%
    \else%
      \setlength{\unitlength}{\unitlength * \real{\svgscale}}%
    \fi%
  \else%
    \setlength{\unitlength}{\svgwidth}%
  \fi%
  \global\let\svgwidth\undefined%
  \global\let\svgscale\undefined%
  \makeatother%
  \begin{picture}(1,0.34023204)%
    \lineheight{1}%
    \setlength\tabcolsep{0pt}%
    \put(0,0){\includegraphics[width=\unitlength,page=1]{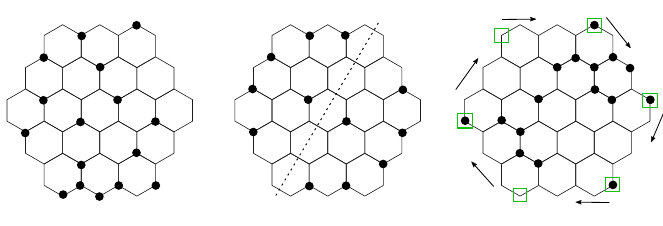}}%
    \put(0.89179638,0.31692843){\color[rgb]{0,0,0}\makebox(0,0)[lt]{\lineheight{1.25}\smash{\begin{tabular}[t]{l}$A$\end{tabular}}}}%
    \put(0.96719592,0.21000242){\color[rgb]{0,0,0}\makebox(0,0)[lt]{\lineheight{1.25}\smash{\begin{tabular}[t]{l}$B$\end{tabular}}}}%
    \put(0.94124329,0.05062862){\color[rgb]{0,0,0}\makebox(0,0)[lt]{\lineheight{1.25}\smash{\begin{tabular}[t]{l}$C$\end{tabular}}}}%
    \put(0.77443368,0.016075){\color[rgb]{0,0,0}\makebox(0,0)[lt]{\lineheight{1.25}\smash{\begin{tabular}[t]{l}$D$\end{tabular}}}}%
    \put(0.68864588,0.12331011){\color[rgb]{0,0,0}\makebox(0,0)[lt]{\lineheight{1.25}\smash{\begin{tabular}[t]{l}$E$\end{tabular}}}}%
    \put(0.72319922,0.27820479){\color[rgb]{0,0,0}\makebox(0,0)[lt]{\lineheight{1.25}\smash{\begin{tabular}[t]{l}$F$\end{tabular}}}}%
  \end{picture}%
\endgroup%

    \caption{\textit{ \emph{(left)} A configuration $C$ with $\rho(C)=1$; \emph{(middle)}: a configuration divided into two sectors; \emph{(right)} Computing the view of robots: corners are highlighted by squares, the reading starts at each corner and proceeds along the side of $G_3$, $LSS(C)=0000010$ $000011111$ $00001001101$ $1011000000$ $001100000$ $0000001$ and it is obtained from $F$.}
}
\label{fig:view}
\end{figure}

\block{View of robots}
In Section~\ref{sec:muset-Gk} we defined the perimeter of $G_k$ as composed of 6 sides, with two consecutive sides sharing one edge. 
If we consider two clockwise consecutive sides, 
we assume that the shared edge belongs to the second side. 
Hence, each edge on the perimeter belongs to a unique side. A \textbf{corner} of $G_k$ is defined as the rightmost vertex on a side of $G_k$. Robots encode the grid $G_k$ starting from a corner of $G_k$ and proceeding clockwise along the side of the grid; then, all the lines parallel to that side are analyzed one by one proceeding along the same direction. When one vertex is encountered, it is encoded as 1 or 0 according to the presence or the absence, respectively, of a robot. This produces a binary string for each starting corner, and then six strings in total are generated. Among such strings, the \textbf{lexicographically smallest string} is denoted as $\LSS(C)$. 
If two strings obtained from different corners are equal, then the configuration is rotational, otherwise it is asymmetric. 
%
%
Hence, the number of generated strings equal to $\LSS(C)$ corresponds to $\rho(C)$. The robot(s) with \textbf{minimum view} is the one with the minimum position in $\LSS(C)$. 
Figure~\ref{fig:view}.right shows the view computed by robots. 

\begin{figure}[t]
   \centering
   \def\svgwidth{0.5\columnwidth}
    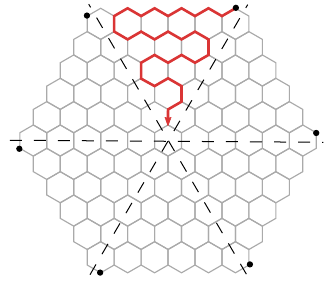
    \caption{The 6 sectors of $G_3$, the corner in each sector, and a \specialpath within one sector.}
\label{fig:special}
\end{figure}

\block{Special-paths}
We have already remarked that $G_k$ can be divided into six sectors. In each sector, we define a \textbf{\specialpath} that starts from each corner of $G_k$ and proceeds as indicated in Figure~\ref{fig:special}. Regarding the number of vertices in a sector and those composing a \specialpath, consider the topmost sector in Figure~\ref{fig:special} and analyze the vertices level by level, starting from level 1: there are $1+3+5+\ldots + 2k-1 = k^2$ vertices in total, while the defined \specialpath leaves one vertex per level (except level 1) untouched, thus having $k^2-(k-1)$ vertices in total. 


\section{A resolving algorithm for $\GMV$ on hexagonal grids}
%
%
According to Theorem~\ref{thm:muset-Gk}, at most $4k$ robots can be arranged in mutual visibility on the vertices of $G_k$. 
As a consequence, $\GMV$ defined on a given $C=(G_k,\lambda)$ can be restricted to $n\le 4k$ robots.
As already remarked, a final configuration  
$C'$ must be provided 
to the algorithm designed to solve $\GMV$. For instance, if $C$ has $4k$ robots, this can be done as follows: let $C'=(G_k,\lambda')$ be the configuration such that $\lambda'(v)=1$ iff $v\in X_k$, where $X_k$ is the $\mu$-set of $G_k$ as defined in Section~\ref{sec:muset-Gk}. 
%

It is worth noting that solving $\GMV$ on $G_k$ by using $X_k$ to define $C'$ resembles the \textbf{pattern formation problem}, where robots belonging to an initial configuration $C$ are required to arrange themselves to form a configuration $F$ (i.e., the pattern) which is provided as input to robots. In this context, it is well-known that ``$\rho(C)$ divides $\rho(F)$'' is a necessary condition for solving the pattern formation problem (e.g., see~\cite{CDDN23}). Since 
$\rho(C')=2$, the existence of only $X_k$ as a $\mu$-set for $G_k$ would restrict the analysis of the problem 
to input instances $C$ such that $\rho(C)\in \{1,2\}$. 


\block{Our Approach} 
We assume that each input configuration $C=(G_k,\lambda)$ contains $n$ robots, with $k\ge 4$ and $12 \le n \le 4k$ (these limitations for $k$ and $n$ are imposed to avoid cases defined by small instances which, as is often the case, would require
specific approaches). Given $C$, we then assume there exists a configuration $F=(G_{k'},\lambda')$ where $\lambda'$ indicates the presence of \textbf{target vertices} instead of robots, and such that all the following conditions hold:  
%
\begin{enumerate}
    \item $k'=\lceil n/4 \rceil$;
    \item $\lambda'$ with at most two targets per line, and at least one target on the perimeter; 
    \item $\rho(C)$ divides $\rho(F)$. \label{necessaryCond}
\end{enumerate}

According to these assumptions, \emph{we design $\algo$ such that, in a finite number of \LCM cycles, it transforms $C$ into a configuration $C'$ having robots disposed as in $F$}. 
%
By using this general approach, if $\algo$ take as input $C=(G_k,\lambda)$ with $n=4k$ robots and $\rho(C) \le 2$, then $X_k$ can be used for defining $F$ as described above; otherwise, if $C$ contains $n\in \{4k-1, 4k-2,4k-3\}$ robots, a corresponding subset of $X_k$ with $n$ elements can be used. So far it is still an open problem to find a $\mu$-set for $G_k$ with symmetricity greater than two. 
But if this problem will be solved in the future, $\algo$ can be used to solve configurations with $\rho(C)>2$.

\subsection{Description of the algorithm}
We present here a high-level description of the algorithm $\algo$ that solves the $\GMV$ problem for fully synchronous robots endowed with chirality and moving on a finite grid $G_k$. Notice that $\algo$ is designed according to the assumptions described in the previous paragraph.
By using the methodology provided in~\cite{CDN21a}, the problem $\GMV$ is divided into a set of sub-problems that are simple enough to be thought of as ``tasks'' to be performed by (a subset of) robots.

The first sub-problem is the ``Placement of guards'', in which $\algo$ selects $\rho(C)$ robots and places each of them on different and symmetric corners of $G_k$. As robots are disoriented (only sharing chirality), the positioning of these robots allows the creation of a common reference system used by robots in the successive stages of the algorithm. For this reason, these special robots are called \textbf{guards}, and each of them will be usually denoted as $r_g$. 
By exploiting chirality, the position of the guards allows robots to identify and enumerate two main directions, hereafter called rows and columns, among the three available in $G_k$. Then, each guard does not move until the final stage of the algorithm. 
The ``Placement of guards'' needs to be decomposed into three simple tasks named $T_{1a}, T_{1b}$, and $T_{1c}$. 

Given the common reference system formed by the guards, all robots can agree on the embedding of the pattern $F$ in $G_k$: robots identify the center of $F$ with $c$ and place the $\rho(F)$ corners of $F$ with the maximum view on the column in which the guards $r_g$ reside and closest to $r_g$. Note that since robots in $C$ are synchronous, irrespective of the algorithm operating on $C$, the center $c$ of the configuration is invariant. 

Task $T_2$ solves the ``Distribute robots on rows'' sub-problem. In particular, $\algo$ first exploits the common reference system formed by the guards to let robots agree on the embedding of the pattern $F$ in $G_k$. Then, $\algo$ moves the robots in each sector along columns to obtain the suitable number of robots for each row according to the targets defined by $F$. During this task, the algorithm avoids occupying any corner of $G_k$ equivalent to the one on which a guard resides (thus preserving the reference system). These vertices are hereafter called \textbf{forbidden}. By doing so, the system of reference is preserved. 

Task $T_3$ solves the ``Moving toward targets'' sub-problem, where all robots but the guards move along rows toward their final target. The moved robots either reach the target or stop one step away from it if their final destination is at a forbidden vertex (there are at most five of such robots as there are at most five forbidden vertices). 

Finally, in task $T_4$, all non-guard robots not on target reach their final destination, and all the guards move away from the corner of $G_k$ toward their final target. In this way, the  ``Pattern finalization'' sub-problem is solved.

\smallskip
In what follows, we detail each of these tasks. 

\block{Task $T_1$} During this task, $\rho(C)$ robots called guards move to occupy a corner of the grid $G_k$. This task is divided into three sub-tasks based on the number of robots occupying the corners of $G_k$. Let $\RP$ be the number of robots on the perimeter, and let $\RC$ be the number of robots on the corners of $G_k$.
Task $T_{1a}$ starts when the following precondition holds: 
$$ \pre_{1a}  \equiv [\RP =0].$$
All robots $r$ such that $D(r)$ is maximum, and of minimum view in case of ties, are selected for moving (for symmetrical reasons, exactly $\rho(C)$ robots are selected). The planned move is the following: $m_{1a}\equiv$ \textit{``each selected robot moves toward a closest vertex belonging to a side of $G_k$, chosen clockwise in case of ties''}. 
%
At the end of $T_{1a}$, $\rho(C)$ robots are on the perimeter of $G_k$. Notice that, when $\rho(C)>1$ each moved robot will be on a distinct side of the grid. 

Task $T_{1b}$ is activated when the following precondition holds:
$$ \pre_{1b}  \equiv [\RP \ge \rho(C) \wedge \RC=0].$$ 
In this case, there are at least $\rho(C)$ robots on the perimeter of $G_k$ but none on corners. Then,
the algorithm selects, on each side, the robot closest to a corner, those with the minimum view in case of ties. Notice that $\rho(C)$ robots are selected.
The planned move is $m_{1b}\equiv $ \textit{``each selected robot moves toward the closest corner of $G_k$, chosen clockwise in case of ties''}.
Finished $T_{1b}$, exactly $\rho(C)$ robots occupy a corner of $G_k$.

Task $T_{1c}$ is activated when the following precondition holds:
$$ \pre_{1c}  \equiv [\RC > \rho(C)].$$
In this case, $\RC - \rho(C)$ robots are moved away from the corners along \specialpaths. The movements are provided by Procedure \MoveAlongSpecialpath\ (cf. Algorithm~\ref{alg:serpentone}). This algorithm uses some additional definitions: a \specialpath is said \textbf{occupied} if its corner (i.e., the first vertex of the path) is occupied; a \specialpath is said to be \textbf{fully-occupied} if  each of its vertices is occupied; 
if $P$ is an occupied \specialpath, but not fully-occupied, then the 
\textbf{\specialsubpath} of $P$ is the maximal subpath that starts at the corner of $P$ and is composed of occupied vertices only. 
Finally, the number of fully-occupied \specialpaths\ is denoted by $\FO$. Note that, due to the maximum number of robots, there cannot be more than one fully-occupied \specialpath.

\begin{algorithm}[t]
\caption{\MoveAlongSpecialpath}\label{alg:serpentone}
\begin{algorithmic}[1]
\begin{small}
 \REQUIRE a configuration $C=(G_k,\lambda)$
     	
    \IF {$\FO=0$}\label{l:nonFully}
        
        \STATE Let $S$ be the set of occupied \specialpaths whose first robot has the minimum view.
        \STATE \textbf{move}: all the robots on a \specialsubpath and not on a \specialpath of $S$ move toward the neighbor vertex along the \specialpath.
      \ENDIF 
      
      \IF {$\FO=1$ }\label{l:Fully}
      \STATE Let $I$ be the fully-occupied  \specialpath
      \STATE \textbf{move}: all the robots on a \specialsubpath of an occupied \specialpath different from $I$ move toward the neighbor vertex along the \specialpath
       \ENDIF
\end{small}
\end{algorithmic} 
\end{algorithm}

At line~\ref{l:nonFully}, the algorithm checks if there are no fully-occupied \specialpaths. 
%
%
The $\rho(C)$ robots occupying corners and having the minimum view are elected as 
guards. 
The move is designed to empty all the other corners of $G_k$ except those elected as guards. In each occupied \specialpaths $P$, except those occupied by guards, the robots on the \specialsubpath of $P$ 
move forward along the \specialpath.
At line~\ref{l:Fully}, there is precisely one fully-occupied \specialpath, say $I$. 
%
%
Therefore, robots on that fully-occupied \specialpath are kept still. Concerning any other occupied \specialpaths $P$ different from $I$, the robots on the \specialsubpath of $P$ move forward along the \specialpath. 
Notice that a configuration where exactly one corner of $G_k$ in each sector is occupied 
is obtained in a single \LCM cycle. 
%

\block{Task $T_2$} This task moves the suitable number of robots for each row according to the pattern $F$. This is the most complex task and it needs several auxiliary concepts.

\begin{figure}[t]
   \centering
   \def\svgwidth{1\columnwidth}
    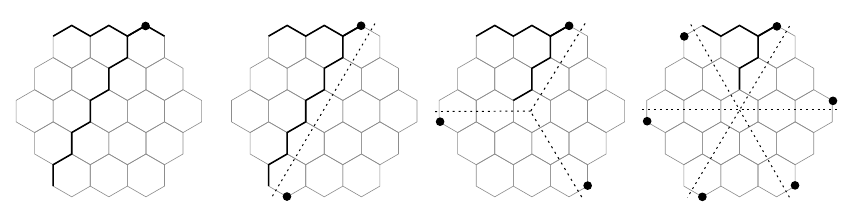
    \caption{\emph{%
    Visualization of the subdivision into c-sectors when $\rho(C)=1,2,3,6$, respectively. In each case, the guard and the row/column with index 0 are emphasized.}
}
\label{fig:sectors_cases}
\end{figure}

When this task starts, we are sure that $T_1$ is concluded and hence the initial configuration $C$ has been transformed so that there are exactly $\rho(C)$ guards positioned on corners. The position of guards allows robots to agree on how to subdivide $G_k$ into $\rho(C)$ 
equivalent sub-configurations hereafter called \textbf{configuration-sectors} (c-sectors, for short). For defining c-sectors, robots use the lines passing through the center of $G_k$ and cutting in half the edge connecting the vertex occupied by the guard robot and the first vertex of the next side (see Figure~\ref{fig:sectors_cases}). There are $\rho(C)$ c-sectors.

Robots can now agree on a \textbf{global reference system} valid in each c-sector. To this end, they can identify two lines among the three available to be used as rows and columns:
a \textbf{row} is any line parallel to the side of $G_k$ in which the guard robot resides (and these lines are assumed as h-lines), and a \textbf{column} is any line obtained by a $60^{\circ}$ counter-clockwise rotation of a h-line (these lines are assumed as r-lines, see Figure~\ref{fig:sectors_cases}). 
%
Moreover, rows and columns can be enumerated within each c-sector as follows: the guard occupies the vertex having coordinates $(0,0,0)$ while all other vertices are labeled as explained in Section 4 by using the coordinates of the guard and the h-lines, r-lines and, as a consequence, l-lines defined in each c-sector.
%
%
Given the common reference system formed by the guards, all robots agree on the embedding of the pattern $F=(G_{k'},\lambda')$ into $G_k$, as follows: 
\begin{definition}\label{def:embedding}
Let $C=(G_k,\lambda)$ be the configuration created by $\algo$ at the end of Task $T_1$, and let $F=(G_{k'},\lambda')$ be the target configuration. The \textbf{\emph{embedding}} of $F$ into $C$ is given by identifying the center of $F$ with the center of $C$ and by superimposing the edges of $F$ into the edges of $C$. Moreover, given a guard, $r_g$ on a side $L$ of $C$, the side $L'$ of $F$ parallel to $L$ and closest to it is the one such that its rightmost target minimizes the distance with $r_g$ and is of minimum view. 
\end{definition}

This embedding allows the algorithm to detect the number of robots to be moved on each row as required by $F$. In particular, $\algo$ identifies $M$ as the number of rows in a c-sector, $t_h$ as the number of targets on row $h$, $(t_1, t_2,\ldots, t_M)$ as the vector of the number of targets, and $(\overline{r_1},\overline{r_2},\ldots, \overline{r_M})$ as the number of robots on each of the rows. 
For each row $h$, the algorithm computes the number of exceeding robots above and below $h$ to the number of targets, to determine the number of robots that need to leave row $h$. Given a row $h$, let $R_{up}(h)=\sum_{i=1}^{h-1} \overline{r_i}$, and let $R_{down}(h)=\sum_{i=h+1}^{M}\overline{r_i}$. Accordingly, let $T_{up}(h)=\sum_{i=1}^{h-1}{t_i}$ and $T_{down}(h) = \sum_{i=h+1}^{M} {t_i}$. 
Concerning the number of targets, given a row $h$, let $\ERup(h)$ be the number of exceeding robots above $h$, $h$ included, and let $\ERdown(h)$ be the number of exceeding robots below $h$, $h$ included. Formally, $\ERup(h) = (R_{up}(h) + \overline{r_h}) \dotdiv (T_{up}(h) + t_h)$, and $\ERdown(h) = (R_{down}(h) + \overline{r_h}) \dotdiv  (T_{down}(h) + t_h)$  where for $a,b$ integers the operator $a \dotdiv b$ is defined as $a \dotdiv b=0$ if $a<b$ and $a \dotdiv b=a - b$, otherwise. 

Let $\MR_{down}(h) = \overline{r_h} - (\overline{r_h} \dotdiv \ERup(h))$ be the number of robots to move downward from row $h$ and let $\MR_{up}(h)= \overline{r_h} − (\overline {r_h} \dotdiv \ERdown(h))$ be the number of robots to move upward from row $h$. Task $T_2$ can be activated only when the following precondition holds:
$$ \pre_2  \equiv [\RC=\rho(C) \wedge \exists \enspace h \in \{1,2,\dots,M\} : \ERdown(h)\neq 0 \lor \ERup(h)\neq 0].$$
Informally, the precondition identifies all the configurations with $\rho(C)$ guards on corners of $G_k$ and at least one row having an excess of robots.

\begin{figure}[h]
   \centering
   \def\svgwidth{0.35\columnwidth}
    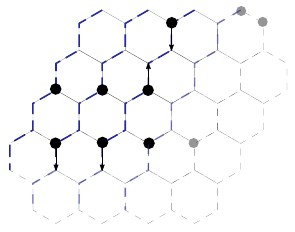
    \caption{\small Robot's movements along columns described in task $T_2$. Black circles represent robots, arrows represent the direction of movements, and bold blue dashed lines show columns.
}
\label{fig:movements}
\end{figure}

%
The move planned in this task is defined in Procedure \MoveAlongColumns\ (cf. Algorithm~\ref{alg:moveRobots}), which is based on the following rationale.  
The procedure moves non-guard robots along columns in each c-sector, where they operate independently and concurrently. Figure~\ref{fig:movements}, shows columns highlighted by bold dashed blue lines. Arrows show the robot's direction of movement. Note that columns do not share any vertex. 

\begin{figure}[h]
   \centering
   \def\svgwidth{1\columnwidth}
     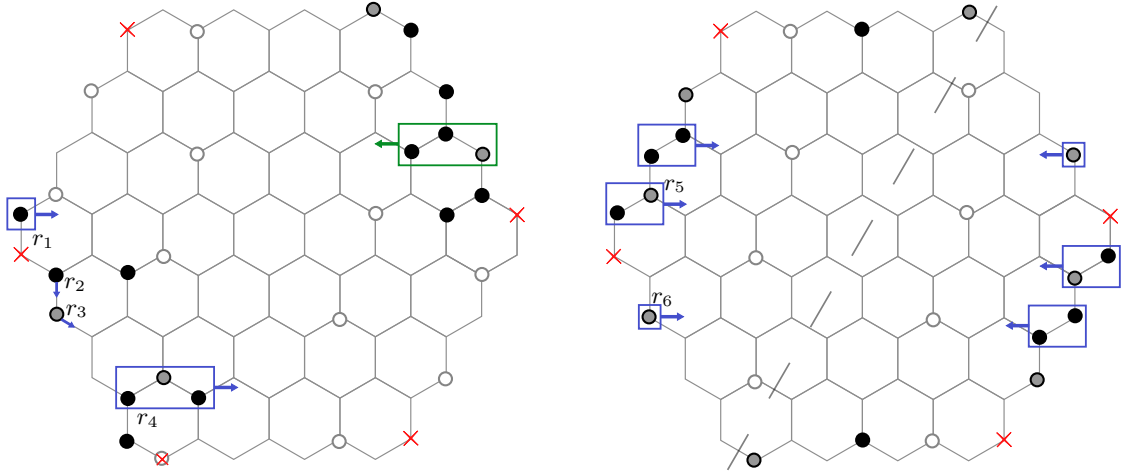
    \caption{\small Two configurations in $G_5$ with all blocked robots. \emph{(left)}: $\rho(C)=1$, \emph{(right)} $\rho(C)=2$. White circles represent targets in $F$, black circles represent robots, and grey circles represent robots already on target. Crosses mark forbidden vertices while rectangles enclose right-shifting paths (left-shifting paths resp.).} 
\label{fig:blocked}
\end{figure}

Since columns have different lengths in $G_k$, in some cases, the robot might be unable to proceed along a column. Another reason why a robot should not proceed is when it could occupy a \textbf{forbidden vertex}, that is any corner of $G_k$ equivalent to the one on which a guard resides (to keep the global reference system). Let $\FV$ be the set containing all the forbidden vertices, then $|\FV| = 6-\rho(C)$. Therefore there are $\frac{6-\rho(C)}{\rho(C)}$ in each c-sector.  
%
%
Let $h$ be a row of a c-sector: robots can distinguish the \textbf{leftmost} (\textbf{rightmost}, resp.) vertex on $h$ as the vertex $v=(l,h,r)$ with minimum (maximum, resp.) $r$-coordinate (and therefore distinguish the left and right side of $h$). To identify the robots to move, let $U_h$ be the set of $\MR_{up}(h)$ robots of row $h$ selected from right and let $D_h$ be the set of $\MR_{down}(h)$ robots of row $h$ selected from the left. $U_h$, $D_h$, and forbidden vertices are used to define the concept of blocked robot (cf. Figure~\ref{fig:blocked}):
%
\begin{definition}\label{def:blocked}
Let $r\in U_h$ ($r\in D_h$, resp.) and belonging to a column $c$, then it is \textbf{blocked-up} (\textbf{blocked-down}, resp.) when one of the following 
holds:
\begin{itemize}
    \item at row $h-1$ ($h+1$, resp.) and column $c$ there is a forbidden vertex,
    \item $h$ is the first (last, resp.) row of column $c$ and $h\neq 1$ ($h\neq M$, resp.),
    \item there is a blocked robot at distance one when moving up (down, resp.),
    \item $\rho(C)=1$, $r$ is located at $h(r)=M-1$ and $c(r)=1$, and there exists a robot $r'$ at $h(r')=M$ and $c(r')=1$.
\end{itemize}
%
Finally, $r$ is \textbf{blocked} when it is blocked-up or blocked-down. 
\end{definition}

Figure~\ref{fig:blocked} shows examples of blocked robots. According to Definition~\ref{def:blocked}, $r_1$ is blocked-down because condition one holds: at row $h(r_1)+1$ and column $c(r_1)$ there is a forbidden vertex; $r_4$ is blocked-down because condition four holds, that is $\rho(C)=1$, $r_4$ is located at $h(r_4)=M-1$, $c(r_4)=1$, and there exists a robot $r'$ at $h(r')=M$ and column $c(r')=1$. Robot $r_5$ is blocked-down because condition three holds, indeed there is a blocked robot at a distance of one while moving down. Finally, robot $r_6$ is blocked because condition two of Definition~\ref{def:blocked} holds, that is $h(r_6$) is the last row of column $c(r_6)$ so robot $r_6$ cannot move down any further. 

Let $H = \bigcup_{h} (U_h \cup D_h)$. If $h$ is a row with $D_h \neq \emptyset$, we call \emph{right-shifting path} the longest path on $h$ starting from the leftmost occupied vertex of $h$ and composed of occupied robots only. Symmetrically, if $h$ is a row with $U_h \neq \emptyset$, we call \emph{left-shifting path} the longest path on $h$ starting from the rightmost occupied vertex of $h$ and composed of only occupied robots. $\mathcal{R}$ ($\mathcal{L}$, resp.) contains all the right-shifting (left-shifting, resp.) paths. Figure~\ref{fig:blocked} shows right-shifting (left-shifting, resp.) paths enclosed in blue (green, resp.) rectangles. The arrows show the movement's direction.

\begin{algorithm}[t]
\caption{\MoveAlongColumns}\label{alg:moveRobots}
\begin{algorithmic}[1]

\REQUIRE $C=(G_k,\lambda)$ and $F=(G_{k'},\lambda')$ 
\STATE compute the view \label{l:view}
\STATE detect guard robots and the corresponding c-sectors
\STATE assign the c-sector vertex labeling by using rows and columns
\STATE perform the embedding of $F$ into $C$
\STATE compute  $U_h$ and $D_h$ for each row $h$; compute $H$, $\mathcal{R}$, and $\mathcal{L}$ \label{l:row-column}
\STATE let $h(r)$ and $c(r)$ be the row and column in which $r$ resides 

\IF {all robots in $H$ are blocked} \label{l:all-blocked}
    \IF { $r$ is not a guard and is on a path of $\mathcal{R}$ ($\mathcal{L}$, resp.)} \label{l:r-shifting}
        \STATE $r$ moves to the right (left, resp.) neighbor on $h(r)$
    \ENDIF
\ENDIF

\IF {$r$ is not a guard and $r \in H$} \label{l:belonging}
     \IF {$r$ not blocked}      
      \STATE $r$ move along $c(r)$ according to the membership to $U_{h(r)}$ or $D_{h(r)}$
      \label{l:move-along-column}

        
    \ENDIF
    
\ELSE
     \IF{ $r$ is not a guard and is $AlignedUp$ or $AlignedDown$ with $r'$} \label{l:collisions} 
        \STATE $r$ moves to a neighbour vertex on $h(r)$ that is not forbidden 

     \ENDIF
 \ENDIF
    
\end{algorithmic} 
\end{algorithm}

We can now describe Procedure \MoveAlongColumns. In the initial block of lines~\ref{l:view}--\ref{l:row-column}, the robot $r$ executing the code computes all the necessary data for the subsequent move: the view, guards, and c-sectors, the vertex labeling local to each c-sector, the embedding of $F$ into $C$, the sets $U_h$ or $D_h$ for each row $h$, and the sets $H$, $\mathcal{R}$, $\mathcal{L}$.  
%
%
%
%
Then, at line~\ref{l:all-blocked} the algorithm checks if all the robots selected to move are all blocked according to Definition~\ref{def:blocked}. In the affirmative case, the robot $r$ (along with all the other robots belonging to the same r-shifting or l-shifting path) performs one horizontal movement toward the assigned direction, see Line~\ref{l:r-shifting}. 
This implies that a fraction of blocked robots will change columns and will be able to move in successive cycles. 
At line \ref{l:belonging}, the robot $r$ checks whether it is selected to move. If it is not blocked, at Line~\ref{l:move-along-column}, it moves along the column $c(r)$ in which it resides either upward or downward depending if it belongs to the set $U_h$ or $D_h$. 
It is possible that, even if $r$ does not belong to $U_h\cup D_h$, $r$ must move because it is ``aligned'' along the column with another robot $r'$ that is involved in the move at Line~\ref{l:move-along-column}. Formally, we say that a robot $r$ is $AlignedUp$ with $r'$ when $r$ resides on row $l$, $U_{l+1}=\{r'\}$  and $c(r)=c(r')$. Similarly, $r$ is $AlignedDown$ with $r'$ when $r$ resides on row $l$, $D_{l-1}=\{r'\}$ and $c(r)=c(r')$.
%
%
%
At line \ref{l:collisions}, $r$ checks if it is $AlignedUp$ or $AlignedDown$ with another robot $r'$. If so, to avoid a collision with $r'$, $r$ moves to a non-forbidden neighbor on $l$. Fig~\ref{fig:blocked}.left shows the case of an $AlignedDown$ robot, $r_2$. Robot $r_2$ is selected to move down, but its target at vertex $h(r_2)+1$, $c(r_2)$ is occupied by another robot, $r_3$. To avoid collisions, $r_2$ lands on its target while $r_3$ moves to a neighbor vertex on $h(r_3)$.
%


\block{Task $T_3$} Non-guard robots move along rows toward their final target. Note that, if a robot encounters a forbidden vertex during its movement, then it will stop at a distance of one from it. During the task, each guard robot will stay in place. This task is activated only when task $T_2$ is over, therefore $\pre_3$ holds:
$$ \pre_3  \equiv [\RC=\rho(C) \wedge \forall \textrm{ row } h: (\ERdown(h)=0 \wedge \ERup(h)=0)].$$
%
According to the conclusion of Task $T_2$, in each row $h$ there are at most two embedded targets of $F$ 
(i.e., $t_h\le 2$) and at most two robots (i.e., $\overline{r_h}\le 2$). Moreover $\overline{r_h}= t_h$. Each non-guard robot $r$ determines its \textbf{target vertex} $t(r)$  as follows: the leftmost robot on $h$ is assigned to the leftmost target in $h$, and the rightmost robot on row $h$ (if any) is assigned to the rightmost target in $h$. Consider the shortest path of $r$ toward $t(r)$, and let $v(r)$ be the next vertex toward $t(r)$ along such a path. Then, move $m_3$ is defined as follows: \textit{ if $v(r) \not \in \FV$ then move to $v(r)$}.

%
%




\block{Task $T_4$} This task is executed only when $T_3$ is concluded, hence all robots matched their final target except for the guards and those robots having a forbidden along the path to reach their final target. Note that, when the guard robots start moving from the corners of $G_k$, the common reference system is lost, hence concepts like guards, c-sectors, rows, and columns cannot be used. 

To verify that $T_4$ must be performed, robots try all the possible embeddings  (at most 6) 
to check whether there exist special conditions. Given $C=(G_k,\lambda)$ and $F=(G_{k'},\lambda')$, we say that an embedding of $F$ into $C$ is \textbf{conclusive} if all the following conditions hold:
\begin{enumerate}
\item 
if $U$ denotes the set of all the unmatched robots of $C$, then $|U|\le 6$,
\item 
$U$ can be partitioned into $U_g$ (set of guards) and $U_f$ (set of robots to be moved on forbidden vertices), where $|U_g|=\rho(C)$ and $0\le |U_f| \le |U|-\rho(C)$,
\item 
if $r\in U_g$, then $r$ is either outside $F$ or on its boundary, and the target of $F$ closest to $r$ is unmatched and located on the boundary of $F$,
\item 
if $r\in U_f$, then $r$ is adjacent to a corner of $G_k$, and this corner corresponds to an unmatched target of $F$.
\end{enumerate}
%
As a consequence, this task can be activated only when the precondition
$$
    \pre_4 \equiv [ \textrm{there exists a conclusive embedding of $F$ into $C$} ]
$$ holds.
The corresponding move is called $m_4$ and is simply defined as follows: \textit{ each robot in $U$ moves toward the unmatched target of $F$ specified in the conclusive embedding of $F$ into $C$}.

\block{Task $T_5$} Each robot recognizes that there exists an embedding of $F$ into $C$ such that all robots are matched (i.e., ``$F$ is formed'') and no more movements are required. The precondition is
$$ \pre_5  \equiv [F~is~formed].$$
%
When this precondition is verified, each robot performs the nil move keeping the current position. 
%

\section{Formalization and Correctness}
The algorithm has been designed according to the methodology proposed in~\cite{CDN21a}. Table~\ref{tab:tasks} summarizes the decomposition into tasks for $\algo$. 
During the execution of $\A$, to detect the task to be accomplished, the \textbf{predicate} $$P_i = \pre_i \wedge \neg ( \pre_{i+1} \vee \pre_{i+2} \vee \ldots \vee \pre_5 )$$ is assigned to $T_i$, for each $i$. 
In the $\Compute$ phase, each robot evaluates 
the predicates starting from $P_5$ and proceeding in the reverse order until a true predicate is found. 
In case all predicates $P_5, P_4, \ldots, P_{1b}$ are evaluated false, task $P_{1a}$ is true and $T_{1a}$ is performed. 
%
%
The provided algorithm $\algo$ can be used by each robot in the $\Compute$ phase as follows: 
%
\emph{-- if a robot $r$ executing $\algo$ detects that predicate $P_i$ holds, and if $r$ is involved in the task,  then $r$ simply performs move $m_i$ associated with $T_i$.}

\begin{table}[t]
\centering
\begin{tabular}{|l|l|l|c|}
\hline
\textit{sub-problems} 
& \textit{task} & \textit{precondition} & \textit{transitions} \\ \hline
\multirow{3}{*}{Placement of guards}  & $T_{1a}$ & $\RP=0$  & $T_{1b}$  \\ \cline{2-4}
& $T_{1b}$ & $\RP \ge$ $\rho(C)$ $\wedge$ $\RC=0$ & $T_{1b}$, $T_2$, $T_3$, $T_4$ \\ \cline{2-4} 
& $T_{1c}$ & $\RC > \rho(C)$ & $T_2$, $T_3$, $T_4$, $T_5$ \\ \hline
\begin{tabular}[c]{@{}l@{}}
Distribute robots on rows
\end{tabular} & $T_2$ & 
\begin{tabular}[c]{@{}l@{}}$\RC=\rho(C)$ $\wedge$ $\exists$ $h \in \{1,2,\ldots, M\}: $\\ 
($\ERdown(h)\neq 0$ $\lor$ $\ERup(h)\neq 0$) \end{tabular} & $T_2$, $T_3$, $T_4$, $T_5$   \\ \hline
Moving toward targets & $T_3$ & \begin{tabular}[c]{@{}l@{}}$\RC=\rho(C)$ $\wedge$ $\forall$ $h\in \{1,2,\ldots, M\}$: \\ ($\ERdown(h)=0$ $\wedge$
$\ERup(h)=0$)  
\end{tabular} & $T_3$, $T_4$, $T_5$  \\ \hline
Pattern finalization  & $T_4$ & \begin{tabular}[c]{@{}l@{}} 
%
%
there exists a conclusive embedding \\ of $F$ into $C$
\end{tabular} & $T_4$, $T_5$ \\ \hline
Termination & $T_5$  & $F$ formed & $T_5$ \\ \hline
\end{tabular}
\vspace{1em}
\caption{Phases of the algorithm: the first column reports a summary of the task's goal, the second column reports the task's name, the third column reports the precondition to enter each task, and the last column reports the transitions among tasks.}
\label{tab:tasks}

\end{table}
%
\begin{figure}[t]
\begin{center}
\begin{tikzpicture}[thick, main/.style = {draw, circle, scale=1.1, }, node distance={20mm}] 
\node[main] (5)  {$~T_{2~}$}; 
\node[main] (3)  [left of=5] {$~T_{1b}$}; 
\node[main] (2) [below left of=3] {$~T_{1a}$}; 
\node[main] (7) [below right of=5] {$~T_{4~}$}; 
\node[main] (8) [below of=7] {$~T_{5~}$};
\node[main] (4) [left of=8] {$~T_{1c}$}; 
\node[main] (6) [left of=4] {$~T_{3~}$}; 
\draw[-latex] (2) to [out=120,in=90,looseness=4] (2);
\draw[-latex] (2) -- node[midway, above right, sloped, pos=0] {} (3);
\draw[-latex] (3) to [out=120,in=90,looseness=4] (3);
\draw[-latex] (3) -- node[midway, above right, sloped, pos=0] {} (5);
\draw[-latex] (3) -- node[midway, above right, sloped, pos=0] {} (6);
\draw[-latex] (3) -- node[midway, above right, sloped, pos=0] {} (7);
\draw[-latex] (4) -- node[midway, above right, sloped, pos=0] {} (5);
\draw[-latex] (4) -- node[midway, above right, sloped, pos=0] {} (6);
\draw[-latex] (4) -- node[midway, above right, sloped, pos=0] {} (7);
\draw[-latex] (4) -- node[midway, above right, sloped, red, pos=0] {} (8); 
\draw[-latex] (5) to [out=120,in=90,looseness=4] (5);
\draw[-latex] (5) -- node[midway, above right, sloped, pos=0] {} (6);
\draw[-latex] (5) -- node[midway, above right, sloped, pos=0] {} (7);

\draw[-latex] (6) to [out=180,in=150,looseness=4] (6);
\draw[-latex] (6) -- node[midway, above right, sloped, pos=0] {} (7);
\draw[-latex] (7) -- node[midway, above right, sloped, pos=0] {} (8);

\draw[-latex] (7) to [out=120,in=90,looseness=4] (7);
\draw[-latex] (8) to [out=270,in=300,looseness=4] (8);
\draw [-latex] (5) edge  [out=0,in=45, looseness=1.1]                   node {} (8);
\draw [-latex] (6) edge  [out=315,in=225, looseness=0.6]                   node {} (8);
\end{tikzpicture}
\end{center}
\caption{Transition graph.} 
\label{fig:transitions}
\end{figure}
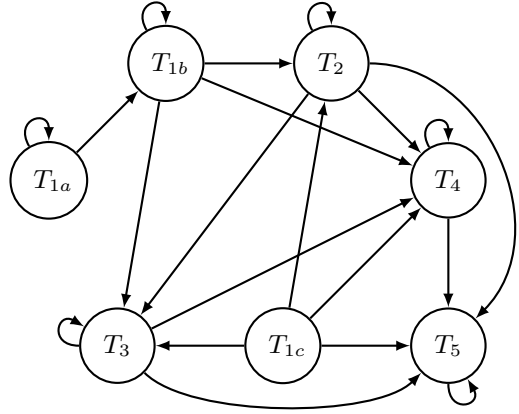

Assume $T_i$ is performed on a current configuration $C$, $\A$ transforms $C$ into $C'$, and $C'$ is assigned the task $T_j$: we say that $\A$ generates a transition from $T_i$ to $T_j$. The set of all possible transitions of $\A$ determines a directed graph called \emph{transition graph} (cf. Figure~\ref{fig:transitions}). 
%
%


We now formally prove that algorithm $\algo$ solves the $\GMV$ problem.  
The correctness of the proposed algorithm can be obtained by proving that all the following properties hold:

\begin{description}
\item[$\h_1$:] 
\textit{for each task $T_i$, the tasks reachable from $T_i$ using any transition are exactly those represented in the transition graph;}
\item[$\h_2$:]
\textit{each transition in the transition graph occurs after a finite number of cycles. This means that the generated configurations can remain in the same task only for a finite number of computational cycles; }

\item[$\h_3$:] 
\textit{the algorithm is collision-free. }

\end{description}

Since these properties must be proved for each transition/move, then in the following we provide a specific lemma for each task. A final theorem will assess the correctness of $\algo$ by making use of all the proved properties $\h_1$--$\h_3$ for each task. 
Note that, regarding property $H_2$, there are no loops involving different tasks in algorithm $\algo$, therefore in the following proofs, we will address only self-loops.

\begin{lemma}\label{lem:corr-T1a}
Let $C$ be a  configuration in $T_{1a}$. From $C$, $\algo$ eventually leads to a configuration belonging to $T_{1b}$.
\end{lemma}

\begin{proof}
During task $T_{1a}$, $\rho(C)$ robots $r_g$ are selected as the robots such that $D(r)$ is maximum.
Let us analyze properties $\h_i$, for $1\le i\le 3$, separately.

\begin{description}
\item[$\h_1$:] 
As the algorithm enters $T_{1a}$, no robots are on the perimeter of $G_k$. The robots selected to move, called $r_g$, go toward the closest side of $G_k$. In successive cycles, the same robots are selected repeatedly, and $\RP=\rho(C)$ and  $\RC=0$  as soon as they reach the perimeter of $G_k$. The precondition $\pre_{1b}$ becomes $\true$ and the algorithm is in $T_{1b}$. The configuration is not in task $T_{1c}$, $T_2$, or $T_3$ since their preconditions require at least one robot on a corner of $G_k$.

\item[$\h_2$:]
At each step, the robots $r_g$ reduce the distance from the perimeter of $G_k$ by one, therefore $\RP$ becomes $\rho(C)$ in a finite number of cycles and the algorithm is not in $T_{1a}$ anymore. 
\item[$\h_3$:]
Since robots $r_g$ are the robots such that $D(r)$ is maximum, they do not meet any other robot while moving toward the perimeter of $G_k$, so no collision can occur, and no multiplicities are created. 
\end{description}
\end{proof}

\begin{lemma}\label{lem:corr-T1b}
Let $C$ be a  configuration in $T_{1b}$. From $C$, $\algo$ eventually leads to a configuration belonging to $T_{1b}$, $T_2$, $T_3$, or $T_4$.
\end{lemma}

\begin{proof}
Task $T_{1b}$ selects $\rho(C)$ robots $r_g$ located on a side of $G_k$ closest to a corner of $G_k$, with the minimum view in case of ties.
Let us analyze properties $\h_i$, for $1\le i\le 3$, separately.

\begin{description}
\item[$\h_1$:] 

Entering task $T_{1b}$, several robots are on the perimeter of $G_k$ but none on corners. Algorithm $\algo$ selects $\rho(C)$ robots to move toward a corner of $G_k$. In the successive cycles, the same robots are repeatedly selected, and when they reach the corners of $G_k$, $\RC=\rho(C)$ and the algorithm is not in $T_{1b}$ anymore.  $\algo$ is either in $T_2$, $T_3$, $T_4$, tasks in which a guard robot is on a corner of $G_k$. 
\item[$\h_2$:]
At each step, the selected $\rho(C)$ robots $r_g$ reduce the distance from a corner of $G_k$ by one step, therefore in a finite number of cycles $\RC$ becomes $\rho(C)$ and the algorithm is not in $T_{1b}$ anymore.

\item[$\h_3$:]
Since the moving robots are those closest to a corner of $G_k$, each of them cannot meet any other robot during the movement toward a corner, therefore no collision can occur and no multiplicities are created. 

\end{description}
\end{proof}

\begin{lemma}\label{lem:corr-T1c}
Let $C$ be a  configuration in $T_{1c}$. From $C$, $\algo$ eventually leads to a configuration belonging to $T_{2}$, $T_3$, $T_4$ or $T_5$.
\end{lemma}

\begin{proof}
During this task, robots move along the $\specialpaths$. 
While entering task $T_{1c}$, there are more than $\rho(C)$ robots on corners of $G_k$. Recall that $\FO$ is the number of fully-occupied $\specialpaths$. It cannot be greater than one as there are  $k^2$ vertices in each $\specialpath$ while robots are at most $4k$ robots and $k\ge4$. Let us analyze properties $\h_i$, for $1\le i\le 3$, separately.

\begin{description}
\item[$\h_1$:] 
When $T_{1c}$ starts, $\RC \ge \rho(C)$. After the move, $\rho(C)$ guard robots are on corners of $G_k$ and $\RC=\rho(C)$. Therefore, the configuration is either in $T_2$, $T_3$, $T_4$ or in $T_5$.
\item[$\h_2$:]
In task $T_{1c}$, all the corners of $G_k$ but $\rho(C)$ are emptied in a robot cycle. 

\item[$\h_3$:]
The \specialpaths are disjoint. During task $T_{1c}$, only robots on special $subpaths$ move along the $\specialpath$. These are the robots on a corner of $G_k$ and the ones in front of it until the first empty vertex. Since robots are synchronous, all these robots move forward by an edge, hence no collision can occur.
\end{description}
\end{proof}

\begin{lemma}\label{lem:corr-T2}
Let $C$ be a  configuration in $T_{2}$. From $C$, $\algo$ eventually leads to a configuration belonging to $T_{3}$, $T_{4}$, or in $T_5$.

\end{lemma}
\begin{proof}
During task $T_2$, $\algo$ moves the robots in each c-sector to place a suitable number of robots for each row, according to the pattern $F$.
Let us analyze properties $\h_i$, for $1\le i\le 3$, separately.

 \begin{description}
\item[$\h_1$:] 
During the movements of robots in task $T_2$, no other corners of $G_k$ get occupied except for the ones already occupied by the guards. Indeed, the set of forbidden vertices is the set of vertices equivalent by rotation to the ones occupied by the guards. A robot $r \in U_h$ ($D_h$, resp.) never moves on a forbidden vertex as it identifies itself as blocked when at row $h-1$ ($h+1$, resp.) and column $c(r)$ there is a forbidden vertex (Condition one of Definition~\ref{def:blocked}). 
Since no guard robot is moved during task $T_2$, the algorithm is never in $T_{1a}$, $T_{1b}$ or $T_{1c}$. At each cycle, the total number of exceeding robots either decreases or remains the same. The last case holds only when all robots are blocked and the shifting movement on the same row is performed. Since robots can stay blocked only for a finite number of cycles, when the number of exceeding robots equals zero for each row, the algorithm is either in $T_3$, $T_4$, or $T_5$.
\item[$\h_2$:]
When the robots in $H$ are not all blocked, the number of exceeding robots decreases at most every two $\LCM$ cycles. When all robots in $H$ are blocked, the robots on right-shifting paths (left-shifting, resp) move on the same row. In this case, the number of exceeding robots per row does not decrease but half of the robots on a right-shifting (left-shifting, resp.) land on a new column and they can proceed in successive cycles. During the algorithm's execution, the shifting movement is performed at most $2k-1$ cycles, corresponding to the length of a side of $G_k$. 

\item[$\h_3$:]
When a robot $r \in H$, it checks if it is blocked, and in the affirmative case, it does not move.  
If all robots $r \in H$ are blocked, then the robots belonging to $\mathcal{R}$ ($\mathcal{L}$, resp) move concurrently to the right (left, resp) on the same row. In this movement, each robot lands on an empty vertex. Indeed there must exist at least one empty vertex in each row, otherwise, not all the robots in $H$ would be blocked.
When a robot $r$ is not blocked and it is in $H$, it moves along its column according to the membership of $U_h$ or $D_h$. Also in this case, $r$ will land on an empty target. Indeed, if a robot $r'$ is on the target of $r$ and it is also selected to move, it will move in the same direction of $r$. Otherwise, if $r'$ it is not selected to move, then it is $AlignedUp$ or $AlignedDowm$. Hence $r'$ moves on an empty neighbor that is not forbidden while $r$ lands on an empty vertex. The fourth condition of Definition~\ref{def:blocked} identifies the condition in which there is a robot $r$ with $c(r)=1$ and $h(r)=M-1$ belonging to $D_h$, in a configuration having $\rho(C)=1$. There is also a second robot $r'$ on the same column of $r$ lying on the last row $M$. Also in this case, $r$ identifies itself as blocked and collisions are avoided.


\end{description}
\end{proof}

\begin{lemma}\label{lem:corr-T3}
Let $C$ be a  configuration in $T_{3}$. From $C$, $\algo$ eventually leads to a configuration belonging to $T_{3}$, $T_{4}$, or in $T_5$.
\end{lemma}

\begin{proof}
Task $T_3$ brings robots toward their final target moving along rows except if a forbidden vertex is along the shortest path of $r$ toward its final target. Then, a robot stops at a distance of one from the forbidden vertex. During this task, the guard robots do not move. 
Let us analyze properties $\h_i$, for $1\le i\le 3$, separately.

\begin{description}
\item[$\h_1$:] 
In task $T_3$, there are $r_l$ robots and $t_l$ targets per row. In each row, robots move toward their final target on their same row. The task continues until $pre_4$ becomes $\true$, and the configuration is in $T_4$ or in $T_5$.
\item[$\h_2$:]
At each $\LCM$ cycle, in each row, robots reduce the distance from their final target by one until either they reach their target or stop at a distance of one from a forbidden vertex. 
\item[$\h_3$:]
There are at most two robots per row and two targets per row. Since the rightmost robot moves toward the rightmost target and the leftmost robot moves toward the leftmost target, collisions are avoided.
\end{description}
\end{proof}

\begin{lemma}\label{lem:corr-T4}
Let $C$ be a  configuration in $T_{4}$. From $C$, $\algo$ eventually leads to a configuration belonging to $T_{5}$.
\end{lemma}

\begin{proof}
In task $T_4$, all robots inside $G_k$ are correctly positioned according to the final pattern except for the guards, and at most five robots on a side of $G_k$ that are at a distance of one from a forbidden vertex. From this configuration, these robots reach their final target while the guards move from a corner of $G_k$ toward their final target.  
Let us analyze properties $\h_i$, for $1\le i\le 3$, separately.

\begin{description}
\item[$\h_1$:] 
The robots located at a distance of one from a forbidden vertex, reach their final target in at most two cycles.
As soon as the guards match their target on $F$, then the pattern is formed, $pre_5$ becomes $\true$ and the configuration is in $T_5$.
\item[$\h_2$:]
At each $\LCM$ cycle, the guards reduce their distance from their final target of one step, and then in a finite number of cycles, they reach their final destination, and the algorithm is not in $T_4$ anymore.
\item[$\h_3$:]

The matched robots perform the nil movement, and the unmatched ones placed at a distance of one from a forbidden vertex move along rows toward their final target. In $\pre_4$, the embedding of $F$ ensures that the target of any guard, in each c-sector, is always on the first row and closer to the guard. That guarantees there is no other robot between a guard and its target. Therefore, no collision can occur as the guard moves toward its final target.
\end{description}
\end{proof}

Concerning complexity, the execution time of $\A$ is determined by the required \LCM cycles. 
Proving the correctness and complexity of $\A$ leads to the following results:

\begin{theorem}\label{thm:main}
If there exists a mutual-visibility set $X$ for $G_k$,
then $\GMV$ can be optimally solved in each configuration $C=(G_k,\lambda)$, $k\ge 4$, composed of $|X|$ synchronous robots endowed with chirality and such that $\rho(C)$ divides $\rho(X)$. 
\end{theorem}
\begin{proof}
Lemmata~\ref{lem:corr-T1a}-\ref{lem:corr-T4} ensure that properties $\h_1$, $\h_2$, and $\h_3$ hold for each task $T_{1a}$, $T_{1b}, \ldots, T_5$. Then, (1) all the transitions are those reported in the transition graph, (2) the generated configurations can remain in the same task only for a finite number of cycles, and (3) the movements of the robots are all collision-free. Lemmata~\ref{lem:corr-T1a}-\ref{lem:corr-T4} show that, from a given task, only subsequent tasks can be reached, or $\pre_5$ eventually holds (and hence $\GMV$ is solved). 
This formally implies that, for each initial configuration $C$ and for each execution $\Ex : C=C(t_0),C(t_1),C(t_2),\ldots$ of $\algo$, there exists a finite time $t_j>0$ such that $C(t_j)$ is similar to the  pattern to be formed in the $\GMV$ problem and $C(t_k) = C(t_j)$ for each time $t_k\ge t_j$. 

The time required by $\algo$ is determined by the number of rounds (i.e., \LCM cycles), as robots are synchronous. From the transition graph in Figure~\ref{fig:transitions}, it follows that there are no cycles except tasks with self-loops. This implies that the total time is trivially bounded by the number of tasks times the maximum number of rounds required for each task. Since each loop in the transition graph can be repeated $O(k)$ times (cf. proofs of Lemmata~\ref{lem:corr-T1a}-\ref{lem:corr-T4}), this leads to a final $\Theta(k)$ time complexity. Indeed, it can be easily observed that each algorithm that solves $\GMV$ requires $\Omega(k)$ rounds. 
\qed
\end{proof}

Theorem~\ref{thm:muset-Gk} shows there exists a mutual-visibility set $X_k$ per $G_k$. From the definition of $X_k$, it is clear that $\rho(X_k)=2$. As a consequence, the following corollary holds.

\begin{corollary}\label{teo:iff}
There exists a time-optimal algorithm that solves $\GMV$ in each 
$C=(G_k,\lambda)$, $k\ge 4$, 
of synchronous robots endowed with chirality and such that $\rho(C)\le 2$. 
\end{corollary}


\section{Conclusion}
In this paper, we gave two main contributions. From a pure graph theoretical point of view, we determined the mutual-visibility number of the hexagonal grid $G_k$ and identified a mutual-visibility set $X_k$ having symmetricity two. Moreover, we have studied the $\GMV$ problem under the \LCM model for synchronous oblivious robots moving on finite hexagonal grids. In this problem, robots located on distinct vertices of an arbitrary graph must reach, in a finite time, a configuration in which they are all in mutual visibility by moving without collisions. We solved $\GMV$ by designing a time-optimal algorithm that exploits the mutual visibility set $X_k$ as a pattern. 
To fully characterize $\GMV$ on hexagonal grids, it is necessary to investigate the existence of $\mu$-sets having symmetricity six. In this direction, we were able to find $\mu$-sets for $G_{3k}$, with $1\le k\le 7$, as shown in Figures from~\ref{fig:G_3-G_6} to~\ref{fig:G_21}. 

Other possible directions of investigation include studying $\GMV$ for $G_k$ under different schedulers (semi-synchronous or asynchronous) and/or with completely disoriented robots (no common chirality). Finally, extending the analysis to general hexagonal grids $G_{k,m,n,}$, with generic integers $k$, $m$, and $n$, would be a challenging research problem.

\section*{Acknowledgments}
S. Cicerone and G. Di Stefano were partially supported by the European Union -
NextGenerationEU under the Italian Ministry of University and Research (MUR) National
Innovation Ecosystem grant ECS00000041 - VITALITY - CUP J97G22000170005, and
by the Italian National Group for Scientific Computation (GNCS-INdAM). A. Di Fonso was partially supported by the Italian Ministry of Economic development (MISE) under the project ``SICURA - Casa intelligente delle tecnologie per la sicurezza", CUP C19C200005200004 - Piano di investimenti per la diffusione della banda ultra larga FSC 2014-2020.
%

\newpage 
\appendix

\newpage
\section{Examples of $\mu$-sets with symmetricity six}

As discussed in the concluding section, we show here optimal mutual visibility sets for $G_2$ and $G_{3k}$, with $1\le k\le 7$, having symmetricity equal to 6. Unfortunately, the $\mu$-sets found for the series of graphs $G_{3k}$ do not present a regularity that allows the generalization of the solution. Finding an optimal mutual visibility set having symmetricity of six remains an interesting graph combinatorial open problem.
\begin{figure}[h]
   \centering
   \def\svgwidth{0.75\columnwidth}
     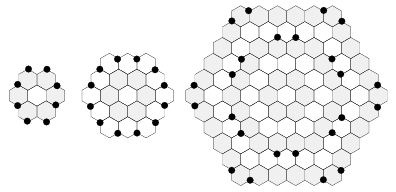
    \caption{\small An optimal mutual visibility sets: from left $G_2$, $G_3$, $G_{6}$. }
\label{fig:G_3-G_6}
\end{figure}

\begin{figure}[h]
   \centering
   \def\svgwidth{0.8\columnwidth}
     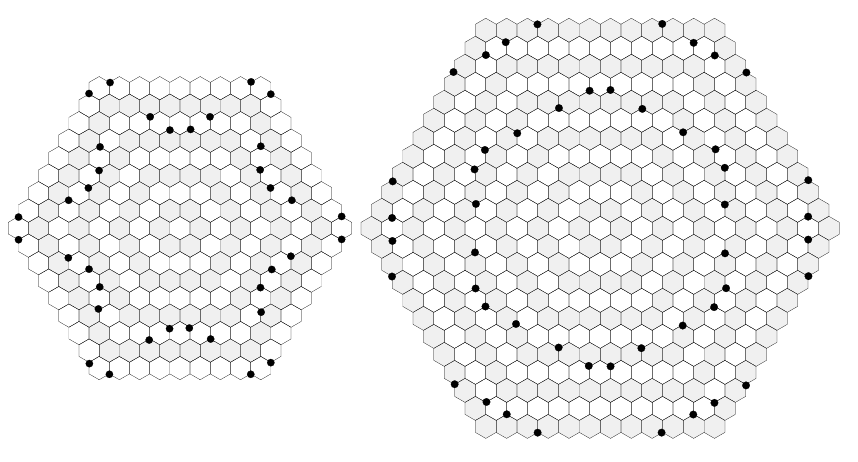
    \caption{\small An optimal mutual visibility sets: left $G_9$, right: $G_{12}$. }
\label{fig:G_9-G_12}
\end{figure}

\begin{figure}[h]
   \centering
   \def\svgwidth{1.1\columnwidth}
     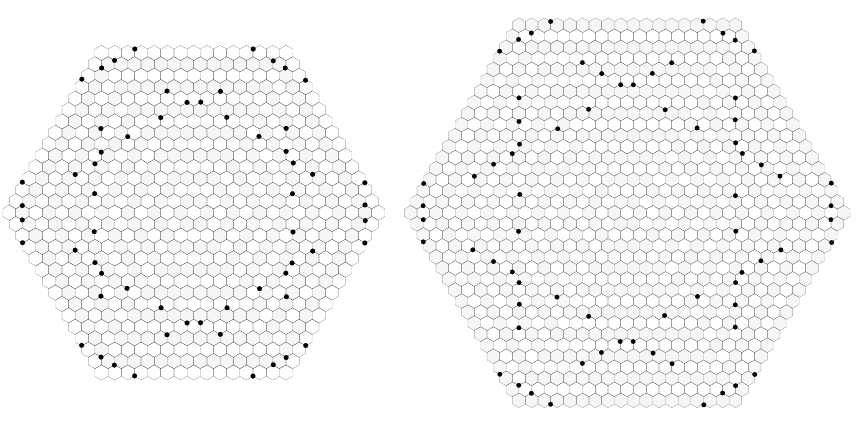
    \caption{\small An optimal mutual visibility sets: left $G_{15}$, right: $G_{18}$. }
\label{fig:G_15-G_18}
\end{figure}

\begin{figure}[h]
   \centering
   \def\svgwidth{0.8\columnwidth}
     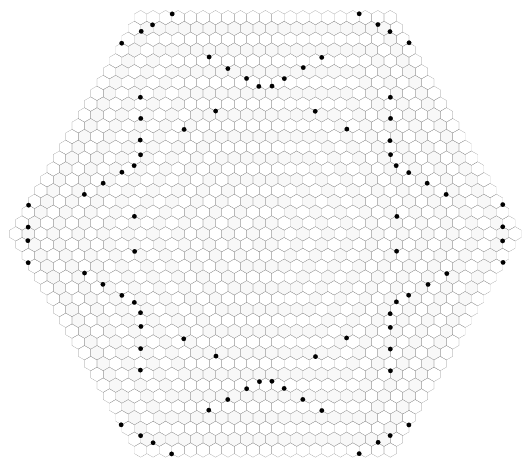
    \caption{\small An optimal mutual visibility set for $G_{21}$. }
\label{fig:G_21}
\end{figure}

\end{document}